\documentclass[aip,reprint,amsmath,amssymb,aps,prd]{revtex4-1}
\usepackage{graphicx}% Include figure files
\usepackage{dcolumn}% Align table columns on decimal point
\usepackage{bm}% bold math
\usepackage{subcaption}
\usepackage{amsmath}
\usepackage{appendix}

\begin{document}
\preprint{AIP}

\author{Ben T. McAllister}
\email{ben.mcallister@uwa.edu.au}
\affiliation{ARC Centre of Excellence for Engineered Quantum Systems, School of Physics, The University of Western Australia, 35 Stirling Highway, Crawley 6009, Western Australia, Australia}
\title{Higher Order Reentrant Post Modes in Cylindrical Cavities}
\author{Yifan Shen}
\affiliation{Kuang Yaming Honors School, Nanjing University, 22 Hankou Rd, Gulou Qu, Nanjing Shi, Jiangsu Sheng, China, 210000}
\author{Graeme Flower}
\affiliation{ARC Centre of Excellence for Engineered Quantum Systems, School of Physics, The University of Western Australia, 35 Stirling Highway, Crawley 6009, Western Australia, Australia}
\author{Stephen R. Parker}
\affiliation{ARC Centre of Excellence for Engineered Quantum Systems, School of Physics, The University of Western Australia, 35 Stirling Highway, Crawley 6009, Western Australia, Australia}
\author{Michael E. Tobar}
\email{michael.tobar@uwa.edu.au}
\affiliation{ARC Centre of Excellence for Engineered Quantum Systems, School of Physics, The University of Western Australia, 35 Stirling Highway, Crawley 6009, Western Australia, Australia}
\date{\today}

\begin{abstract}
Reentrant cavities are microwave resonant devices employed in a number of different areas of physics. They are appealing due to their simple frequency tuning mechanism, which offers large tuning ranges. Reentrant cavities are, in essence, 3D lumped LC circuits consisting of a conducting central post embedded in a resonant cavity. The lowest order reentrant mode (which transforms from the $TM_{010}$ mode) has been extensively studied in past publications. In this work we show the existence of higher order reentrant post modes (which transform from the $TM_{01n}$ mode family). We characterize these new modes in terms of their frequency tuning, filling factors and quality factors, as well as discuss some possible applications of these modes in fundamental physics tests. The appendix contains a comment on a paper related to this work.
\end{abstract}
\maketitle
\section{Introduction}
The cylindrical reentrant cavity is a device that can provide high-Q microwave modes with large tuning ranges. It consists of a metal cavity with a conducting post or ring located centrally within the cavity. The gap between the top of this post or ring and the top of the cavity adjusts the mode frequency and at certain gap spacings traps the electric field within the gap. Such cavity designs have been extensively studied and allow for a standard reentrant mode tuning range on the order of GHz without the need for physically large cavities.~\cite{HighQ,SplitRing,HigherOrder,RCBkgn3,RCBkgnd2,RCBkgnd}

Since this structure was first investigated in connection with the development of klystrons, it has been widely used in the construction of microwave oscillators and particle accelerators\cite{RCOsc,RCParticleAcc}, and is often chosen as a structure in the study of metamaterials~\cite{RCMeta1,RCMeta2}. Some recent work has discussed potential applications in telecommunications~\cite{telecom}, and detection of gravitational waves~\cite{RCgrav,RCgrav2} and dark matter~\cite{RCHalo}. It is an interesting perspective to view the standard reentrant mode as a perturbed $TM_{010}$ mode. The standard reentrant mode transforms into the empty cavity $TM_{010}$ mode as the central post is removed from the cavity~\cite{Rigorous}. In a similar way, there exist higher order reentrant modes, which can be viewed as perturbed $TM_{01n}$ modes.

In this work we unequivocally demonstrate the existence of these higher order reentrant modes and characterize them in terms of their tuning range, and quality factor. Section II presents a theoretical study of these modes based on finite element analysis, and experimental data follows in Section III.
\section{Finite Element Analysis of the Reentrant Cavity}
\begin{figure}[h!]
	\includegraphics[width=0.55\columnwidth]{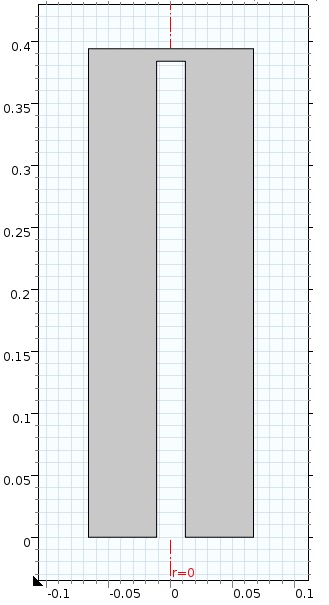}
	\caption{Cylindrical cross-section of the cavity from the COMSOL model in the $r$-$z$ plane (encompassing the origin) in cylindrical coordinates. Grey represents the empty cavity space, while the metallic cavity region is represented by grid-lined white area. The height of cavity is $\sim$0.4 m and the diameter of the inner wall is 0.1337 m. The diameter of the central post is 0.024 m.}
	\label{fig:1}
\end{figure}
\begin{figure*}[t!]
	\includegraphics[width=0.7\textwidth]{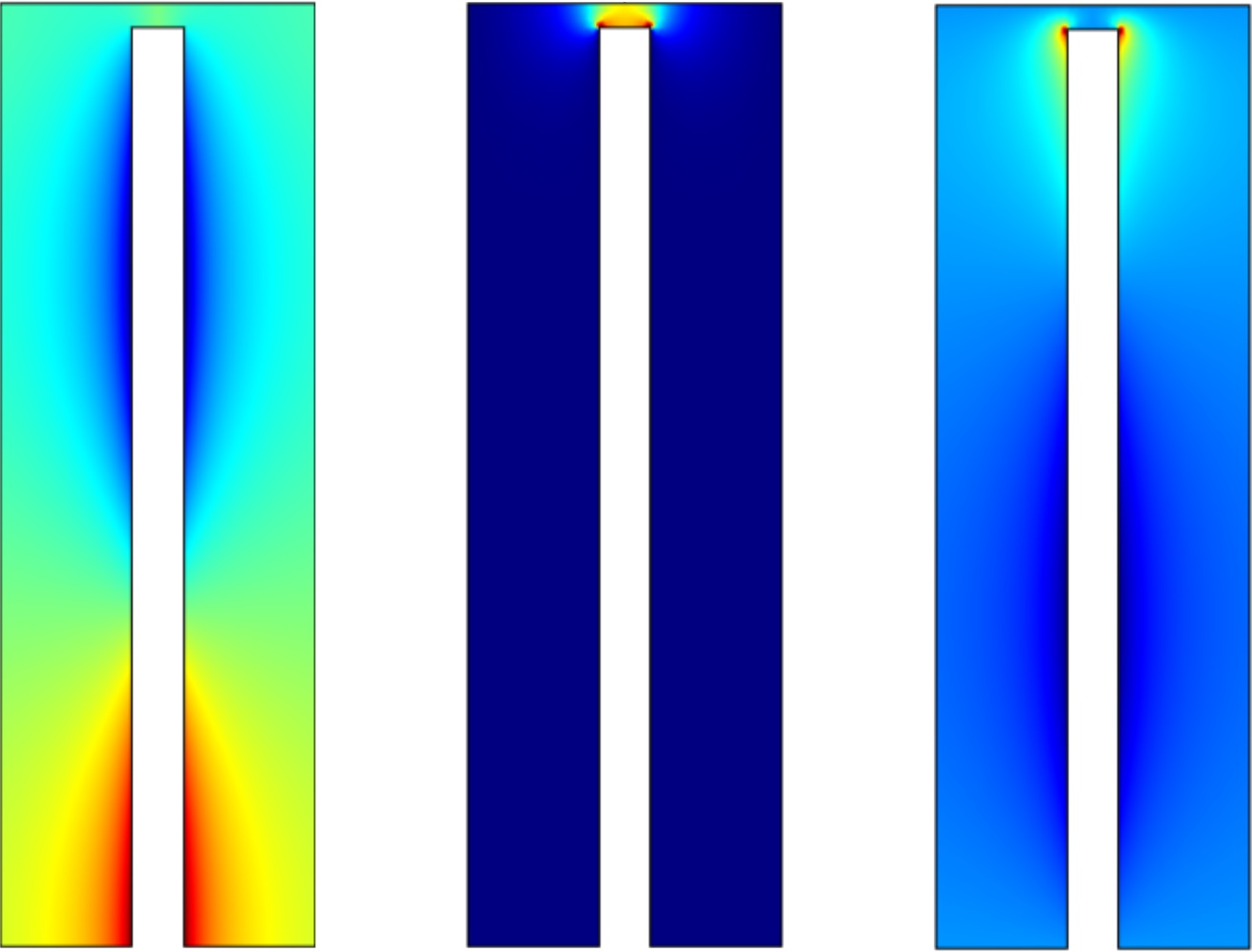}
	\caption{$B_{\phi},~E_z$ and $E_r$ field components for the first higher order reentrant mode are shown from left to right, with a gap size of 1 cm.}
	\label{fig:2}
\end{figure*}
\begin{figure}[b]
	\includegraphics[width=\columnwidth]{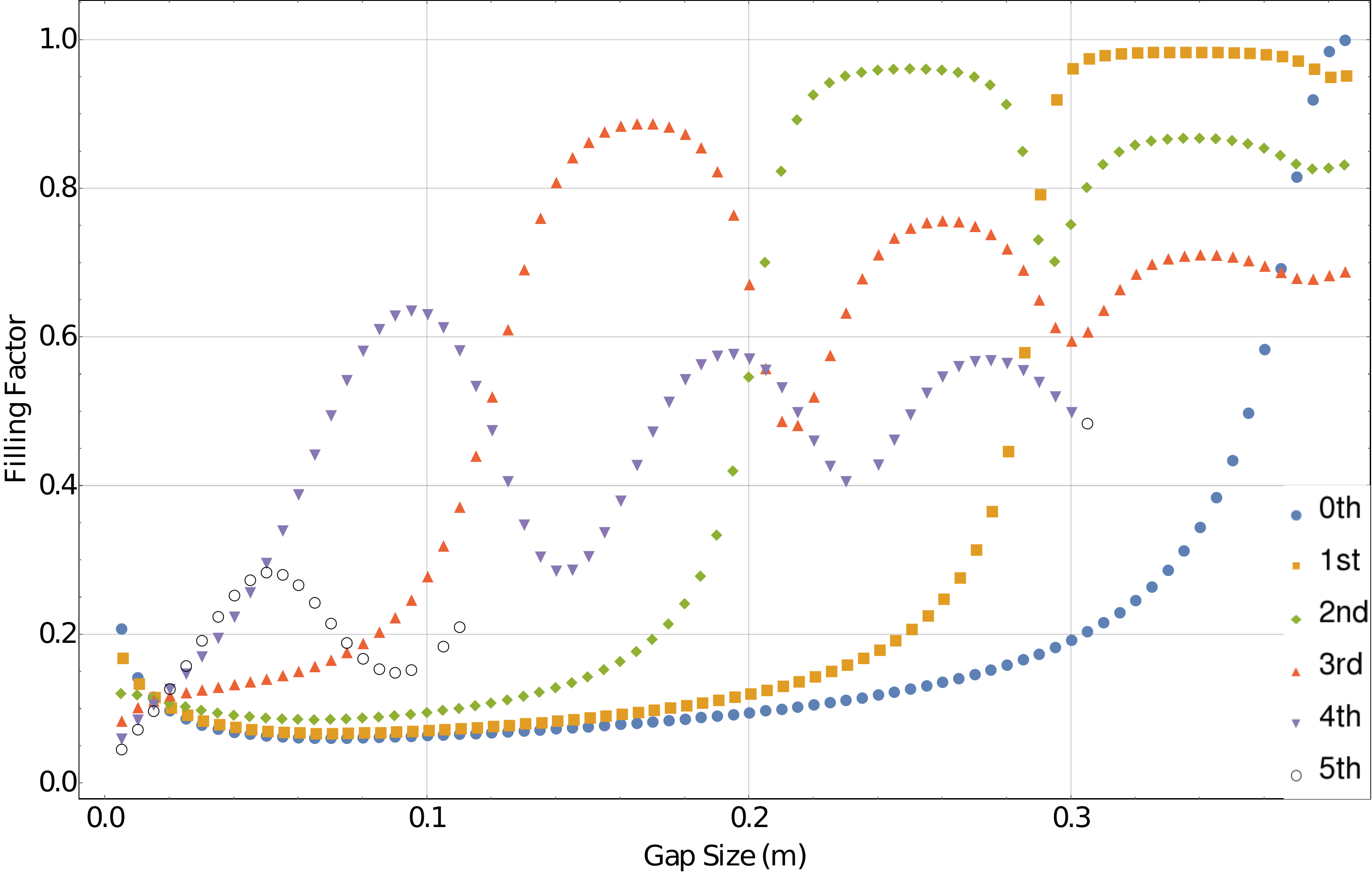}
	\caption{$E_z$ filling factors for multiple reentrant modes as a function of gap size. The modes are represented as detailed in the legend, where the ``0th" order mode corresponds to the fundamental TM$_{010}$-like mode. Points above 2.2 GHz are suppressed.}
	\label{fig:3}
\end{figure}
 \begin{figure*}[t!]
 	\includegraphics[width=\textwidth]{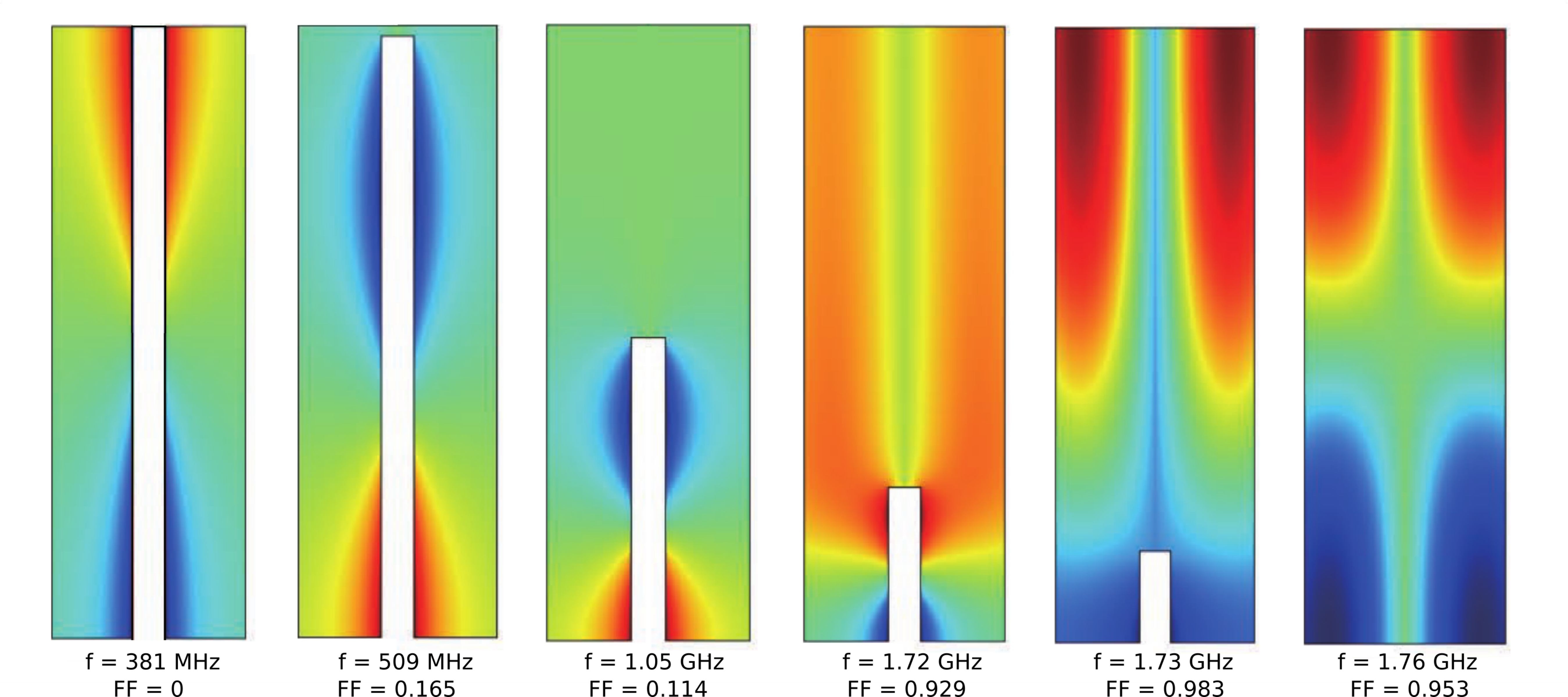}\\
 	\caption{$B_\phi$ component transformation of the 1st higher order reentrant mode. From left to right: the first image represents the corresponding coaxial mode, the second and third images represent the purely reentrant mode. The fourth image shows the transition from the reentrant phase to the `psuedo-TM' phase. The fifth image shows the `pseudo-TM' phase of the 1st higher order mode. The sixth image shows the $TM_{011}$ mode in the cylindrical cavity. Redder regions represent regions of higher positively signed magnetic field, whilst bluer regions represent regions of higher negatively signed magnetic field. The `f' and `FF' values correspond to the frequency and $E_z$ filling factor respectively.\\}
 	\label{fig:4}
 	\includegraphics[width=\textwidth]{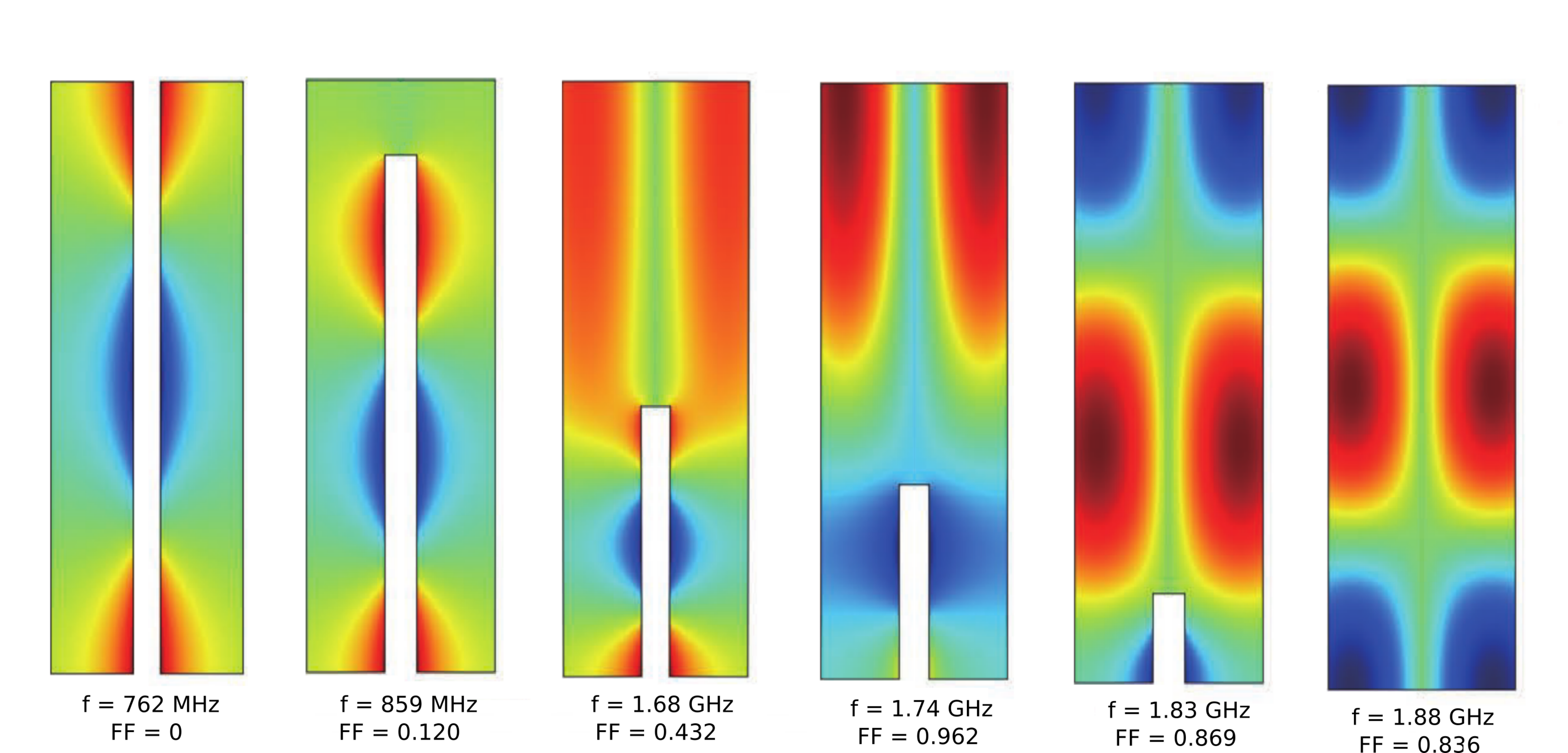}\\
 	\caption{$B_\phi$ component transformation of the 2nd higher order reentrant mode. From left to right: the first image represents the corresponding coaxial mode, the second image represents the purely reentrant mode. The third image shows the transition from the reentrant phase to the `psuedo-TM' phase. The fourth and fifth images show the `pseudo-TM' phase of the 2nd higher order mode, where some of the ``magnetic rings" have moved from the post to the cavity. The sixth image shows the $TM_{012}$ mode in the cylindrical cavity. Redder regions represent regions of higher positively signed magnetic field, whilst bluer regions represent regions of higher negatively signed magnetic field. The `f' and `FF' values correspond to the frequency and $E_z$ filling factor respectively.\\~\\}
 	\label{fig:5}
 \end{figure*}
 ~\\~\\~\\~\\~\\
We employed the COMSOL Multiphysics software to simulate a reentrant cavity structure. Since the single post reentrant cavity is an axisymmetric structure, it sufficed to utilize 2D modelling, with revolution around the z-axis. The structure of interest is shown in fig.~\ref{fig:1}.

We find that multiple higher order modes in the single post reentrant cavity follow similar mode structures to the standard reentrant mode. All of the these modes have the same dominant field components: the axial electric field, $E_z$, which is located in the region directly above the post and below the top of the cavity, the azimuthal magnetic field $B_\phi$, which is separated into several rings surrounding the central post, in alternating phase, and a radial electric field, $E_r$, due to the gradient of $B_\phi$ in the z direction. Fig.~\ref{fig:2} shows the features of these modes.

The main feature of interest in the higher order modes is the azimuthal magnetic field, which alternates in direction between clockwise and counter-clockwise around the post. We name these higher order modes by their axial wave number, or the number of these ``magnetic rings" minus one. Unlike the standard reentrant mode, the higher order modes store their electric energy both in the gap region and cylinder region. The former is determined by the strong $E_z$ field in the central post gap region, and the latter is determined by the radial electric field in the cylinder region which is strongest at the nodes of magnetic field between the ``magnetic rings".

Largely due to the radial electric field, higher order mode filling factors differ from the standard reentrant mode. This can be readily observed when the size of the gap becomes very small. Fig.~\ref{fig:3} shows the $E_z$ filling factor of different modes, defined as
\begin{align}
	FF=\frac{\int dV\left|E_z\right|^2}{\int dV\left|E_c\right|^2}.
	\label{eq:FF}
\end{align}
\begin{figure*}
	\centering
	\begin{subfigure}[t]{0.45\textwidth}
		\includegraphics[width=\textwidth]{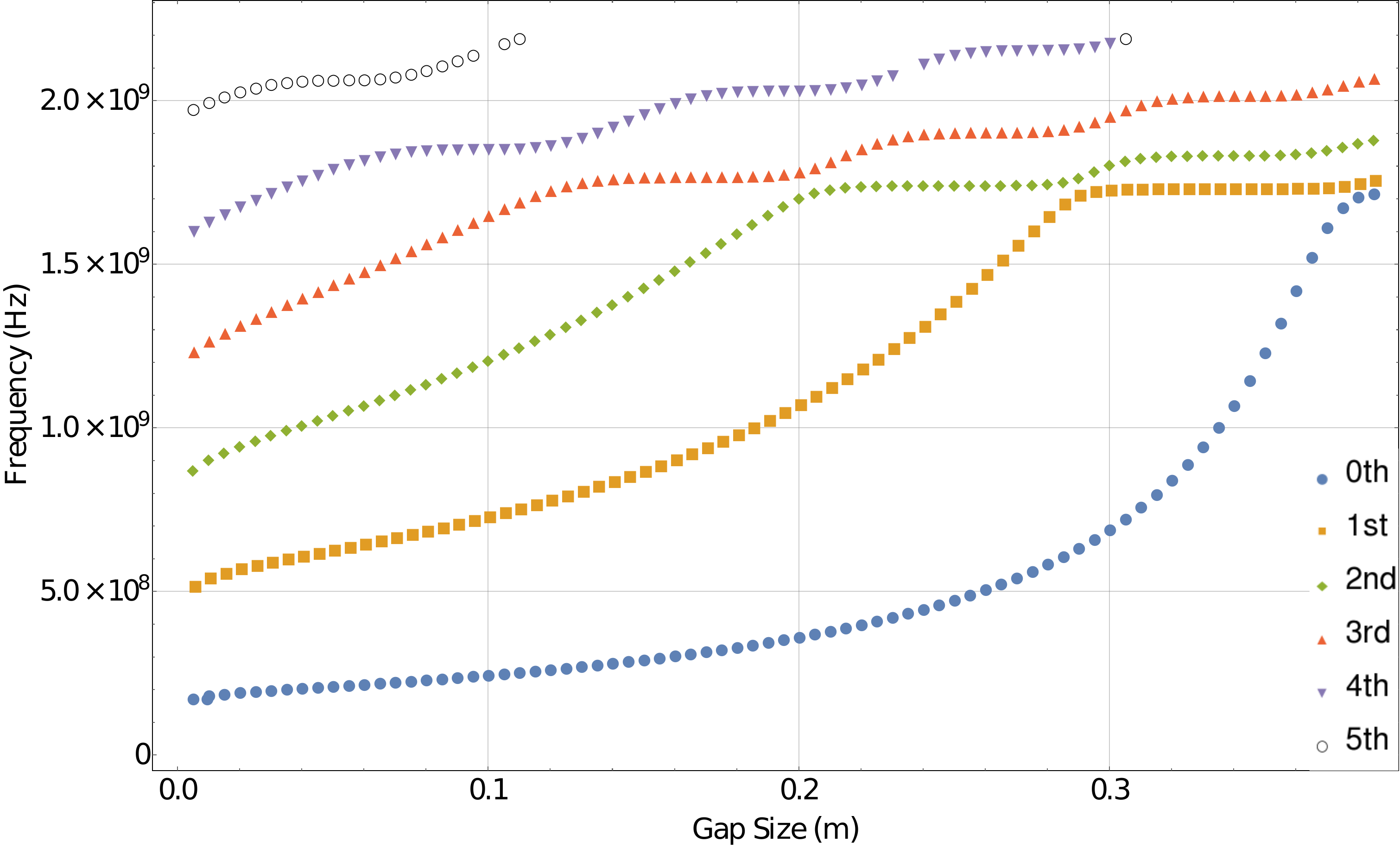}
		\caption{Frequencies of multiple reentrant modes as a function of gap size. The modes are represented as detailed in the legend, where the ``0th" order mode corresponds to the fundamental TM$_{010}$-like mode. Points above 2.2 GHz are suppressed. Simulated data begins with a gap size of 5mm in this run.}
		\label{fig:6}
	\end{subfigure}
	\begin{subfigure}[t]{0.43\textwidth}
		\includegraphics[width=\textwidth]{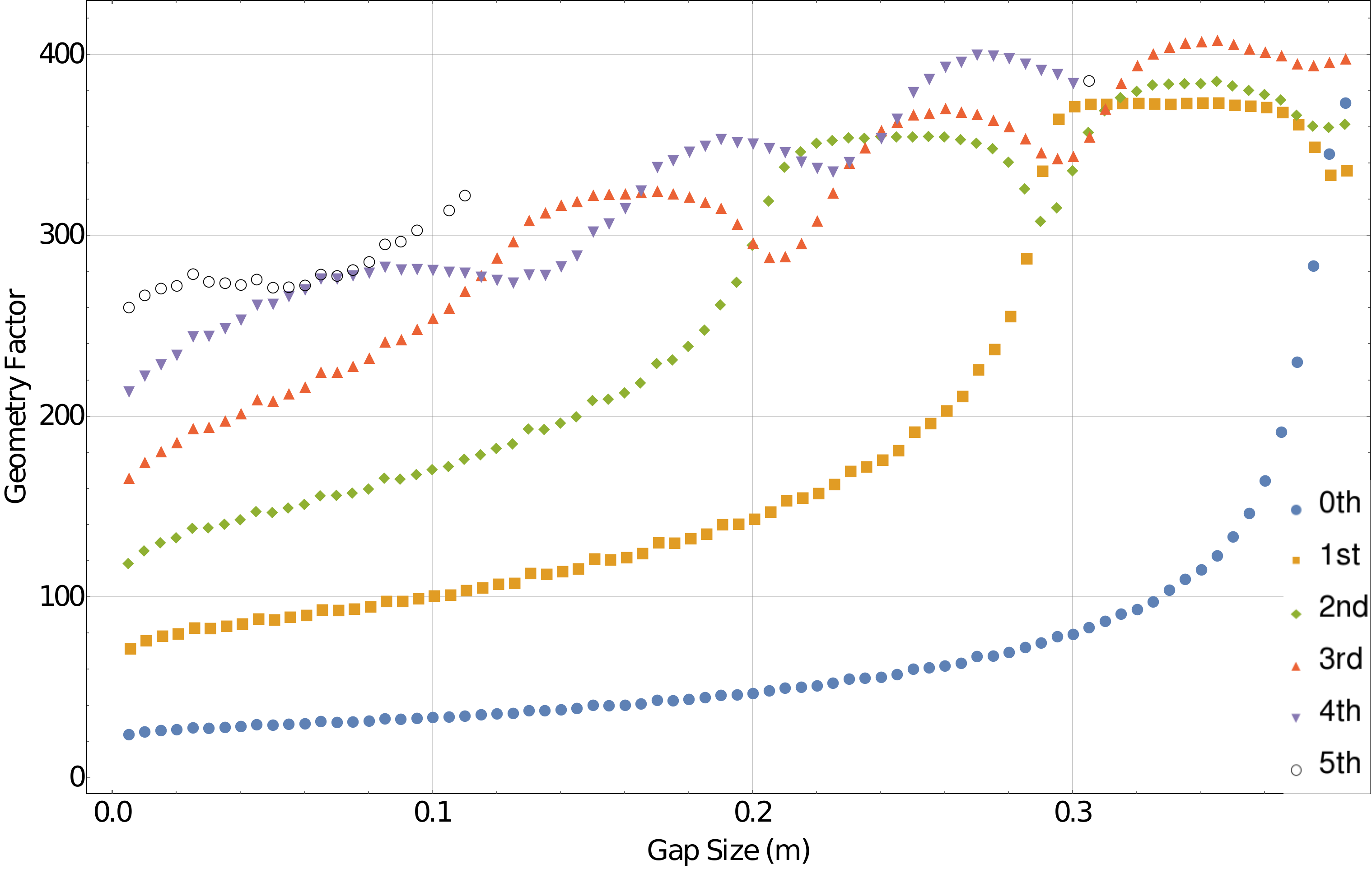}
		\caption{Geometry factors of multiple reentrant modes as a function of gap size. The modes are represented as detailed in the legend, where the ``0th" order mode corresponds to the fundamental TM$_{010}$-like mode. Points above 2.2 GHz are suppressed. Simulated data begins with a gap size of 5mm in this run.}
		\label{fig:7}
	\end{subfigure}
	\caption{}
\end{figure*}
\begin{table*}[t]
	\centering
	\begin{tabular}{|c|c|c|c|c|c|}
		\hline
		Frequency (GHz)&Standard Mode&1st higher order mode&2nd higher order mode&3rd higher order mode&4th higher order mode\\ \hline
		Initial&0&0.381&0.761&1.140&1.520\\ \hline
		Final&1.72&1.71&1.73&1.73&1.84\\ \hline
		Tuning&1.72&1.33&0.967&0.584&0.313 \\
		\hline
	\end{tabular}
	\caption{Tuning data for several reentrant modes. The initial frequencies are the frequencies of the modes when the gap size is zero (coaxial regime) where as the final frequencies are the mode frequencies at the end of each modes large tuning phase, when they begin to transition to an empty cavity mode.}
	\label{tab:1}
\end{table*}
~\\
~\\
~\\
~\\
~\\
Where $E_c$ is the electric field in the cavity. As a consequence of the $E_z$ filling factor of the fundamental $TM_{010}$ mode being unity, we can think of $E_z$ filling factors as normalization of mode field patterns relative to the fundamental. When the gap is of the order of mm, the capacitance of the structure formed by the two planes (the top of the post and the top of the cavity) becomes large, and the electric field in this area becomes stronger. In the standard reentrant mode, with a small $E_r$ electric field component, which itself is near the gap region, the $E_z$ filling factor grows larger at small gaps. However, because the filling factor is related to the integration volume, and due to the existence of larger radial electric field components in the larger non-gap region, the Ez filling factors of higher order modes may first increase as the gap decreases  but eventually approaches zero when the volume of the gap region becomes small.\\
Each higher order mode is slightly different, the first higher order mode reaches a maximum $E_z$ filing factor (about 0.2) when the gap is around 0.0015m, yet other modes' $E_z$ filling factors continually decrease as the gap becomes smaller.
Unlike the standard reentrant mode, these higher order mode frequencies do not tend to zero as the gap size tends to zero, as their effective capacitance does not tend to infinity due again to the radial electric field components. These modes transition to coaxial modes of fixed frequency when the gap size becomes zero~\cite{Coax}.

A common use of a reentrant cavity structure is to utilize the transformation process from empty cavity modes to reentrant modes in order to achieve high frequency tuning ranges. We find that then nth order reentrant mode arises from the transition of a TM01n mode in an empty cylindrical cavity.

Figs.~\ref{fig:4} and~\ref{fig:5} show the transformation process of the 1st and 2nd higher order modes. As we can see in the figures, because $TM_{011}$ and $TM_{012}$ exhibit periodic change in $\phi$ direction magnetic field, the transformation process is complex. As the post moves out of the cavity the top ``magnetic ring" will transform into the upper lobe of the $TM_{01n}$ empty cylinder mode. As the post continues to move the other rings do the same. The higher order modes undergoes multiple transition phases, from coaxial to reentrant, and from reentrant to empty cylinder via a `pseudo-TM' phase where some of the ``magnetic rings" have become empty cavity lobes. Each mode transitions slightly differently, due to the number of ``magnetic rings" and lobes in the empty cavity mode. These transition phases affect frequency tuning.

We have calculated the resonant mode properties as the gap size changes from 0.1mm to $\sim$40 cm. Fig~\ref{fig:6} shows the eigenfrequencies of different reentrant modes. Table~\ref{tab:1} lists tuning ranges of the standard reentrant mode and some high order modes’ first phases. As mentioned, higher order modes go through multiple phases, with the first phase, the purely reentrant phase, exhibiting the largest tuning range. After this point the modes become quite crowded in frequency space, as they are nearly at the frequencies of the corresponding empty cavity TM modes. By monitoring different modes, it is possible tune through different frequency ranges at the same time, which would be beneficial in many fundamental physics tests such as searches for dark matter and gravitational waves.

\begin{figure}[]
	\includegraphics[width=\columnwidth]{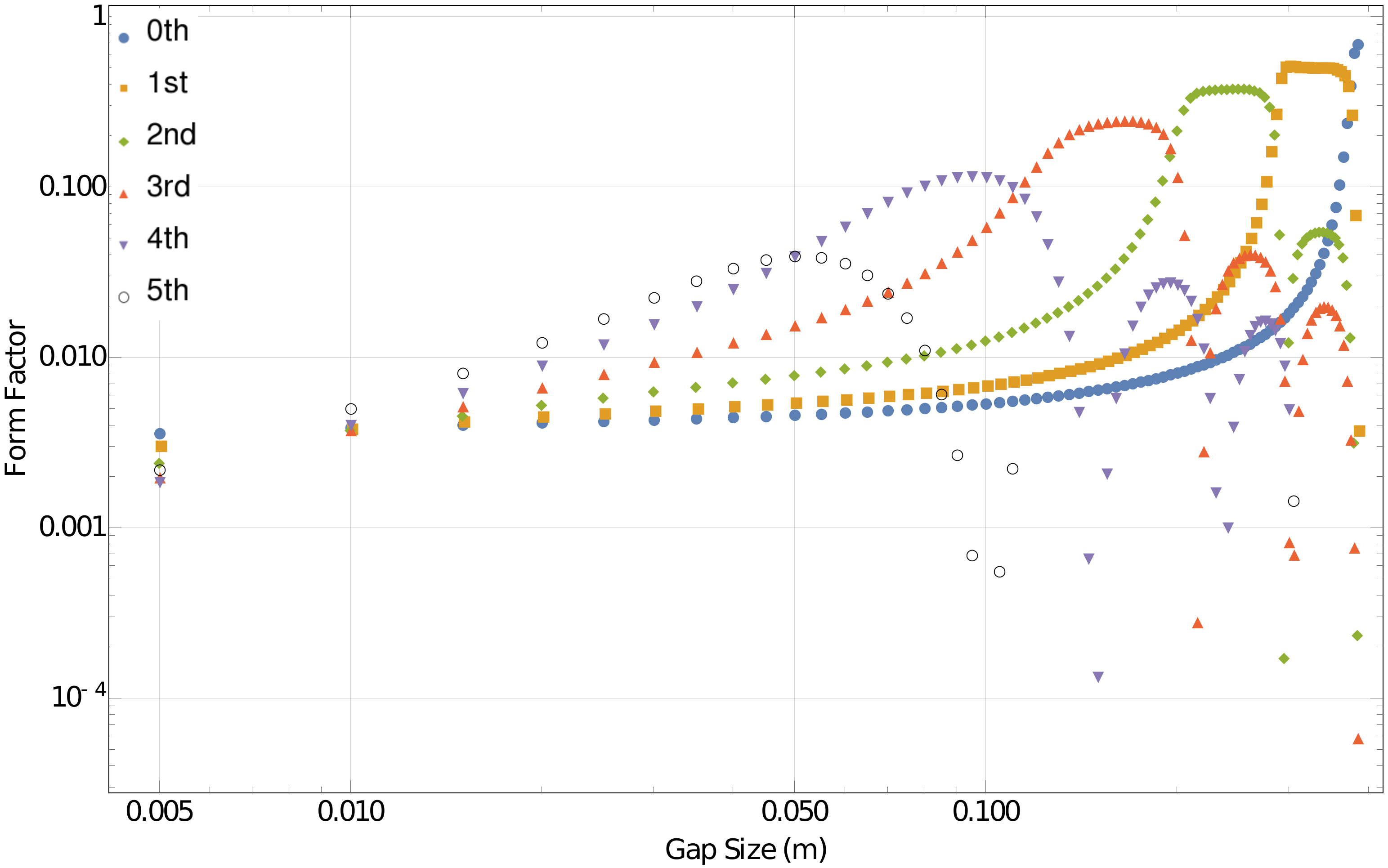}
\caption{Electromagnetic form factors of multiple reentrant modes as a function of gap size. The modes are represented as detailed in the legend, where the ``0th" order mode corresponds to the fundamental TM$_{010}$-like mode. Points above 2.2 GHz are suppressed.}
\label{fig:EMFF}
\end{figure}

\begin{figure}[]
	\includegraphics[width=\columnwidth]{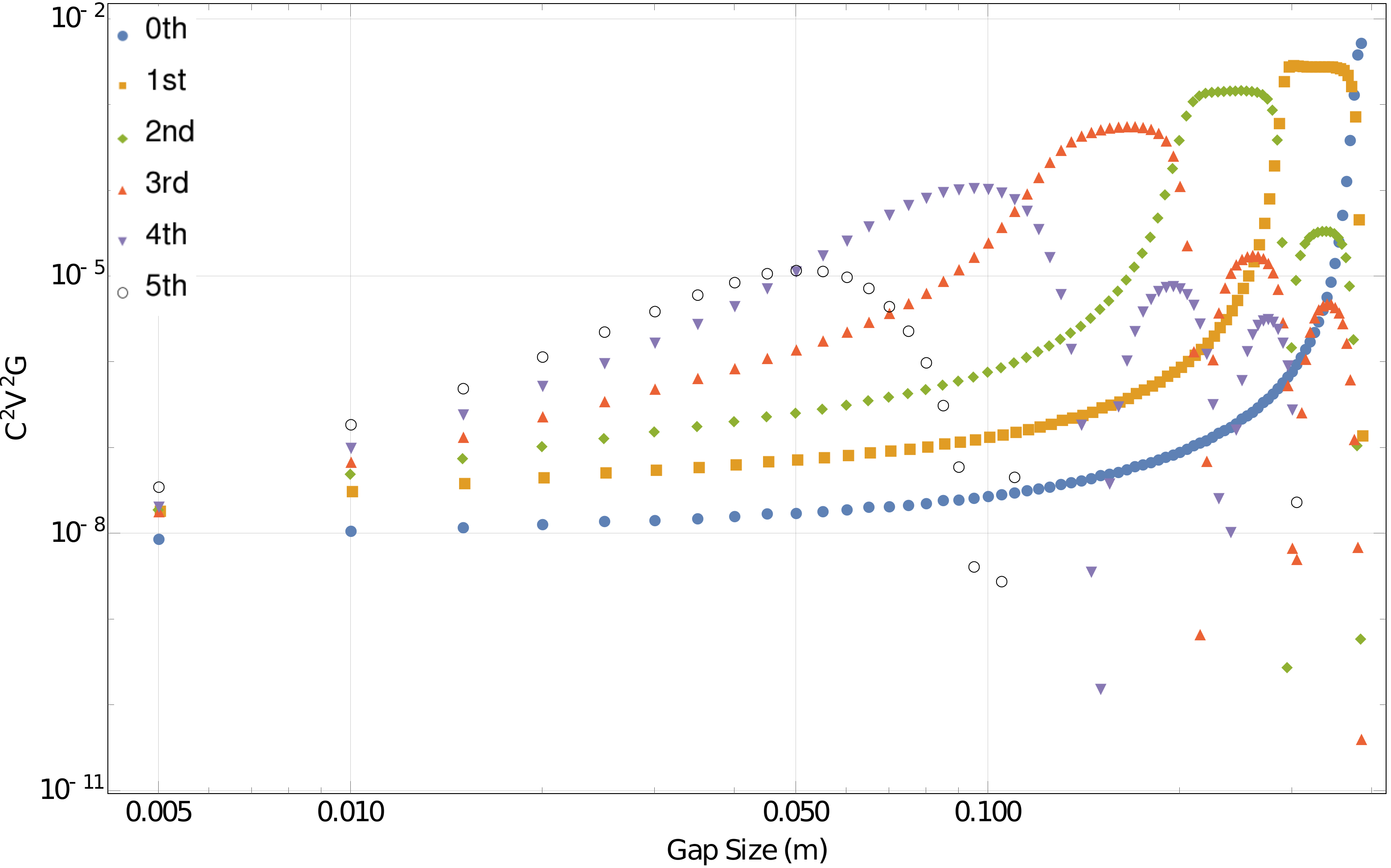}
	\caption{$C^2V^2G$ products of multiple reentrant modes as a function of gap size. The modes are represented as detailed in the legend, where the ``0th" order mode corresponds to the fundamental TM$_{010}$-like mode. Points above 2.2 GHz are suppressed.}
		\label{fig:C2V2G}
	\end{figure}

We have also characterized these modes in terms of their geometry factor, which is directly proportional to mode quality factor
\begin{align}
	\begin{split}	
G&=\frac{\omega\mu_0\int\left|\vec{H}\right|^2dV_c}{\int\left|\vec{H}\right|^2dS_c}\\
Q_{cav}&=\frac{G}{R_s}.
\end{split}
\label{eq:Q}
\end{align}
Where $R_s$ is the surface resistivity of the metal. Fig.~\ref{fig:7} shows the geometry factors as a function of gap size for various reentrant modes. It is obvious that the geometric factor of higher order modes is much higher than that of the standard mode. High-Q modes are highly sought after in fundamental physics tests~\cite{RCgrav,RCgrav2,RCHalo}, which makes these modes promising candidates. One potential application of this class of modes is as a type of receiver known as a haloscope~\cite{Sikivie83Haloscope,Sikivie1985,RCHalo}, to search for axion dark matter~\cite{PQ1977,IpserSikivie1983}. In haloscope searches, one of the most critical parameters is the axion electromagnetic form factor~\cite{McAllisterFormFactor}, which defines the overlap between a cavity mode electromagnetic field, and the electromagnetic field generated by the axion in the presence of an external magnetic field. This factor is entirely mode dependent, and thus a good measure of the suitability of a given mode for axion haloscopes. Previously, electromagnetic form factors and axion sensitivity of the standard, lowest order reentrant mode have been computed~\cite{RCHalo}, fig.~\ref{fig:EMFF} shows the electromagnetic form factors for the first few higher order reentrant modes as a function of gap size. Furthermore, it can be shown that $C^2V^2G$, where C is the form factor, V is the volume of the resonator and G is the geometry factor, should be employed as a figure of merit in axion haloscope design~\cite{DielectricAxion}.C is explicitly, generally defined as
\begin{equation}
	\text{C}=\frac{\left|\int dV_{c}\vec{E_c}\cdot\vec{\hat z}\right|^2}{2~V\int dV_{c}\epsilon_r\mid E_c\mid^2}+\frac{\frac{\omega_a^2}{c^2}\left|\int dV_{c}\frac{r}{2}\vec{B_c}\cdot\vec{\hat\phi}\right|^2}{2~V\int dV_{c}\frac{1}{\mu_r}\mid B_c\mid^2},
\end{equation}
For arbitrary dielectric and magnetic materials. We present $C^2V^2G$ products as a function of gap size in figure~\ref{fig:C2V2G}.
This is interesting, as we can see that the most axion sensitive mode changes as the gap size changes. This could be of interest in axion haloscope searches, as it would open up the possibility of scanning multiple frequency ranges in a single post sweep. It would be possible to begin the search in one range, tracking the lowest order mode, and then move to higher order modes as they become more sensitive, creating a wider feasible scanning range within the same cavity.
\section{Experimental data}
\begin{figure}[]
	\includegraphics[width=0.9\columnwidth]{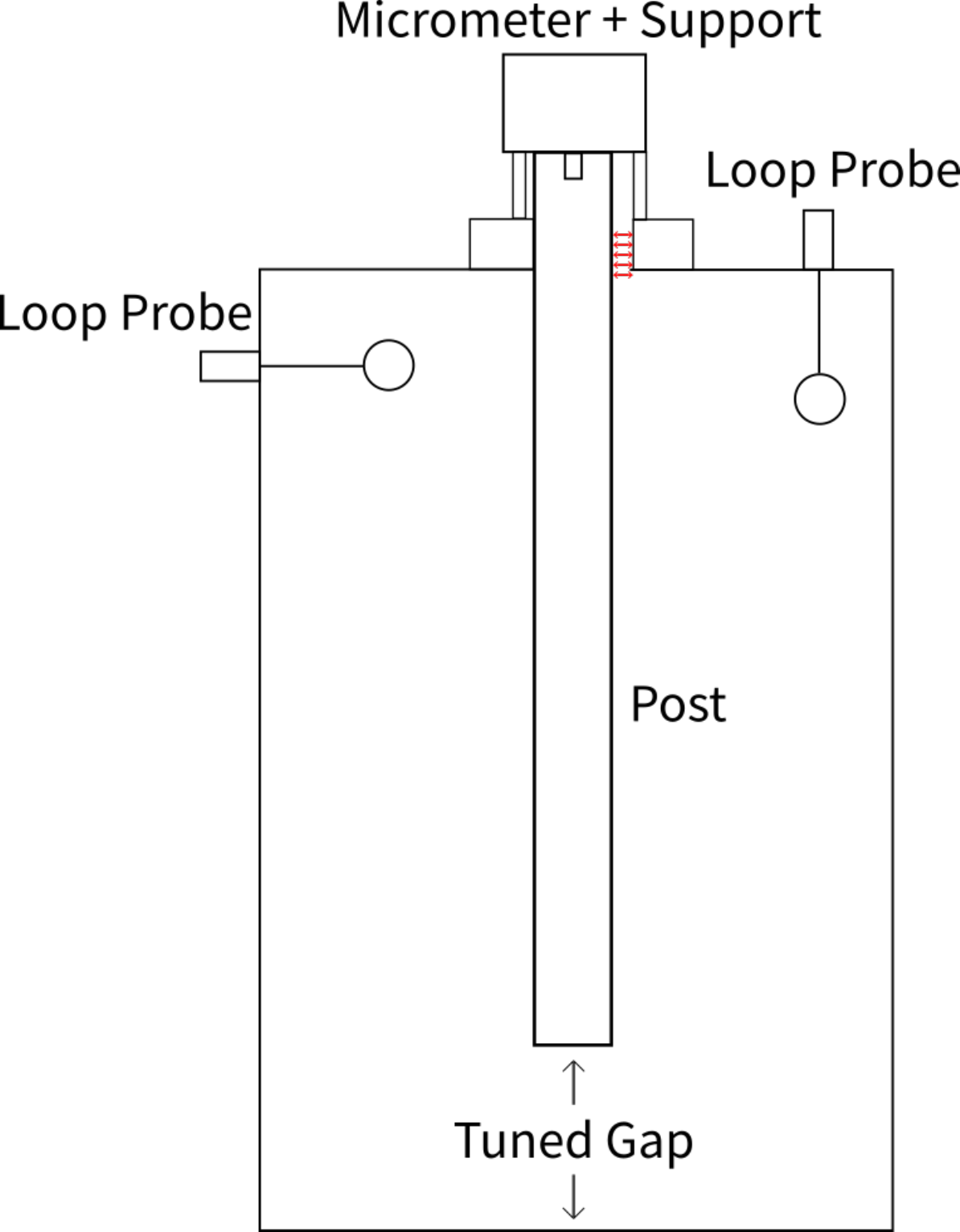}
	\caption{A cross-section of the cavity used for the experiment. The critical components are labelled, whilst the small gap region between the post and the lid referred to in the text is highlighted by the small red arrows. This is the gap which we suggest is responsible for much of the unaccounted for losses. Dimensions are exaggerated for clarity.}
	\label{fig:8}
\end{figure}
\begin{figure}[t!]
	\includegraphics[width=\columnwidth]{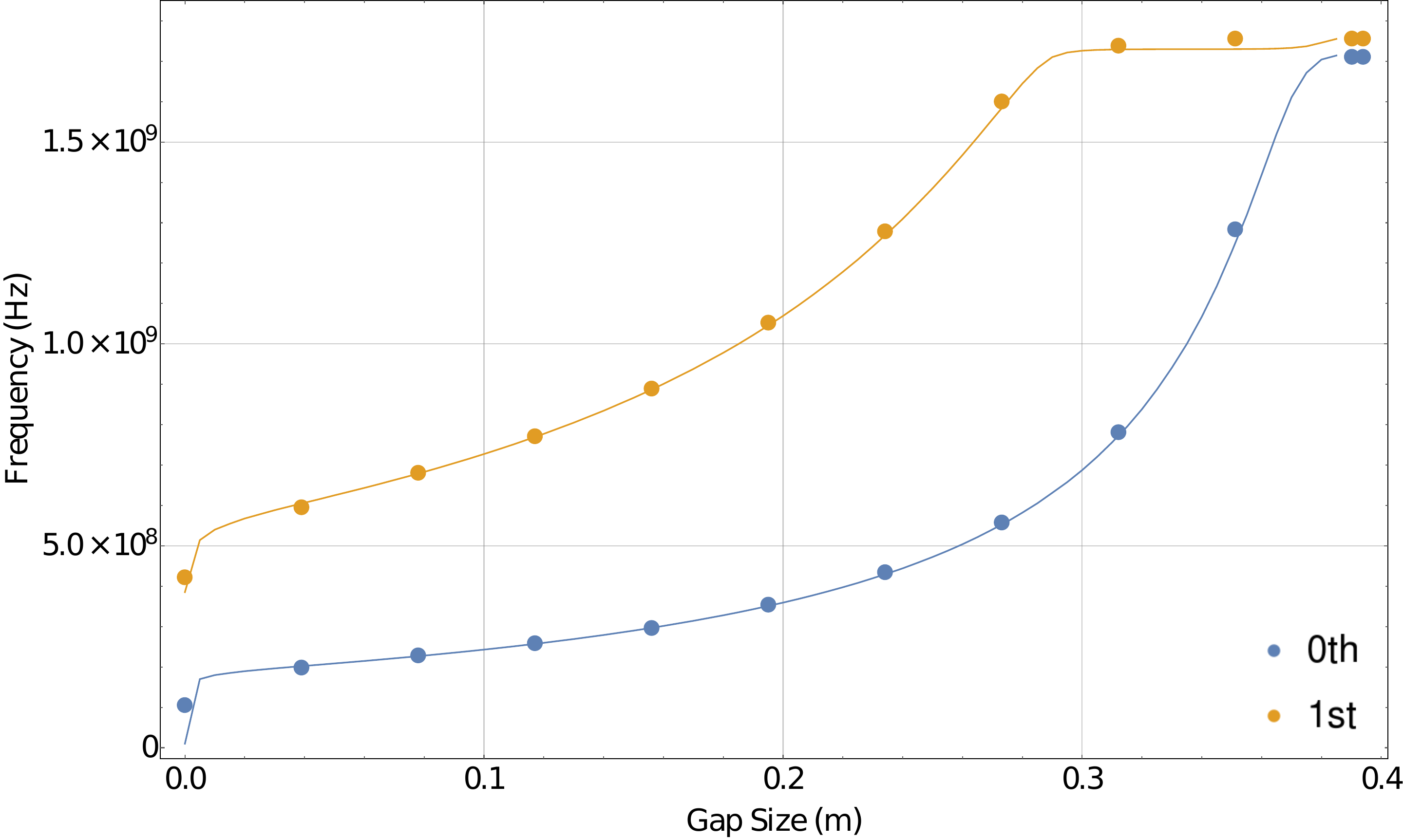}
	\caption{Experimental and theoretical frequency data of the standard and first higher order reentrant mode. The average error is below 3\%. The modes are represented as detailed in the legend, where the ``0th" order mode corresponds to the fundamental TM$_{010}$-like mode.}
	\label{fig:9}
\end{figure}
\begin{figure*}
	\begin{subfigure}[t]{0.45\textwidth}
	\includegraphics[width=\textwidth]{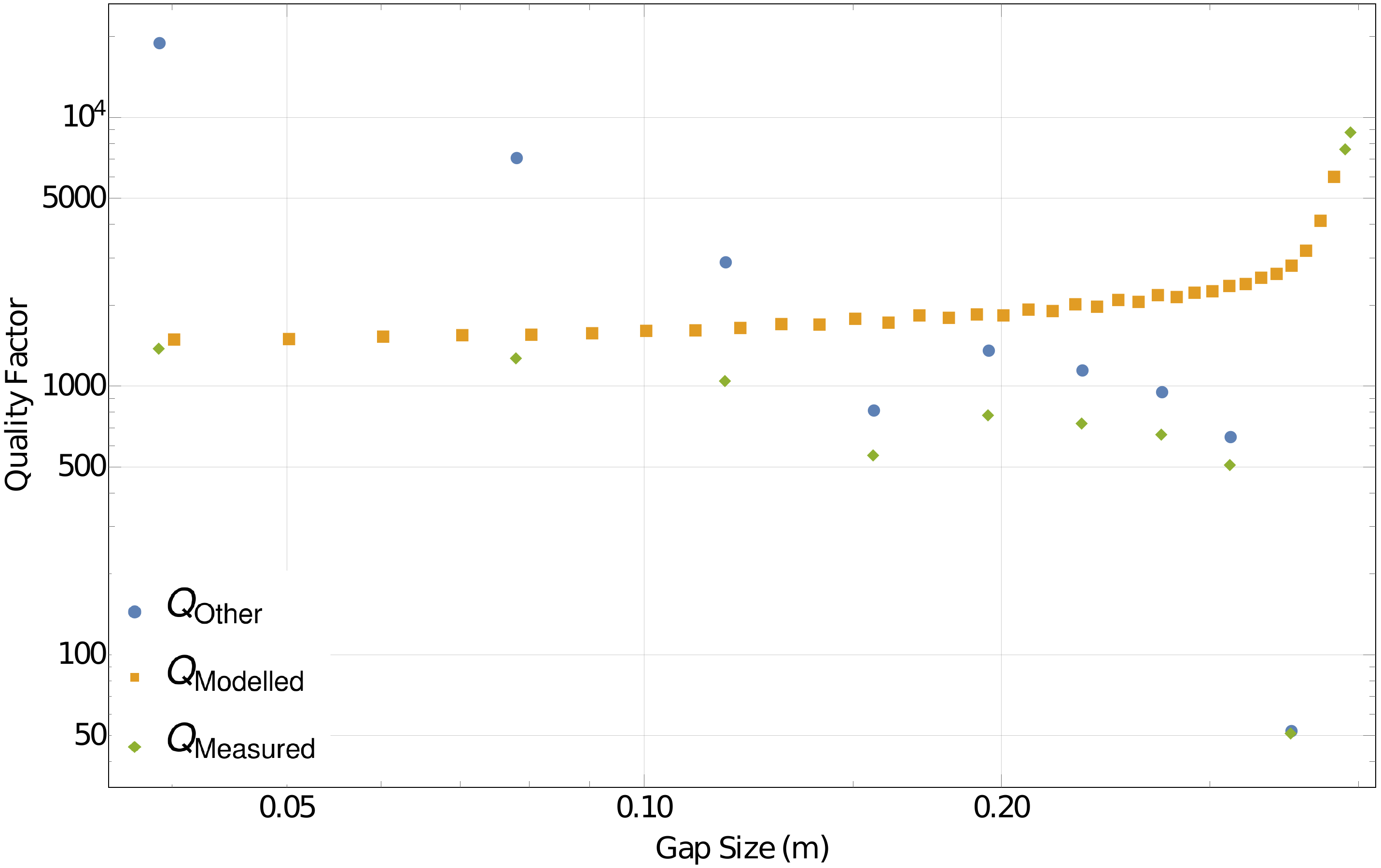}
	\label{fig:Q1}
	\caption{}
\end{subfigure}
\begin{subfigure}[t]{0.455\textwidth}
	\includegraphics[width=\textwidth]{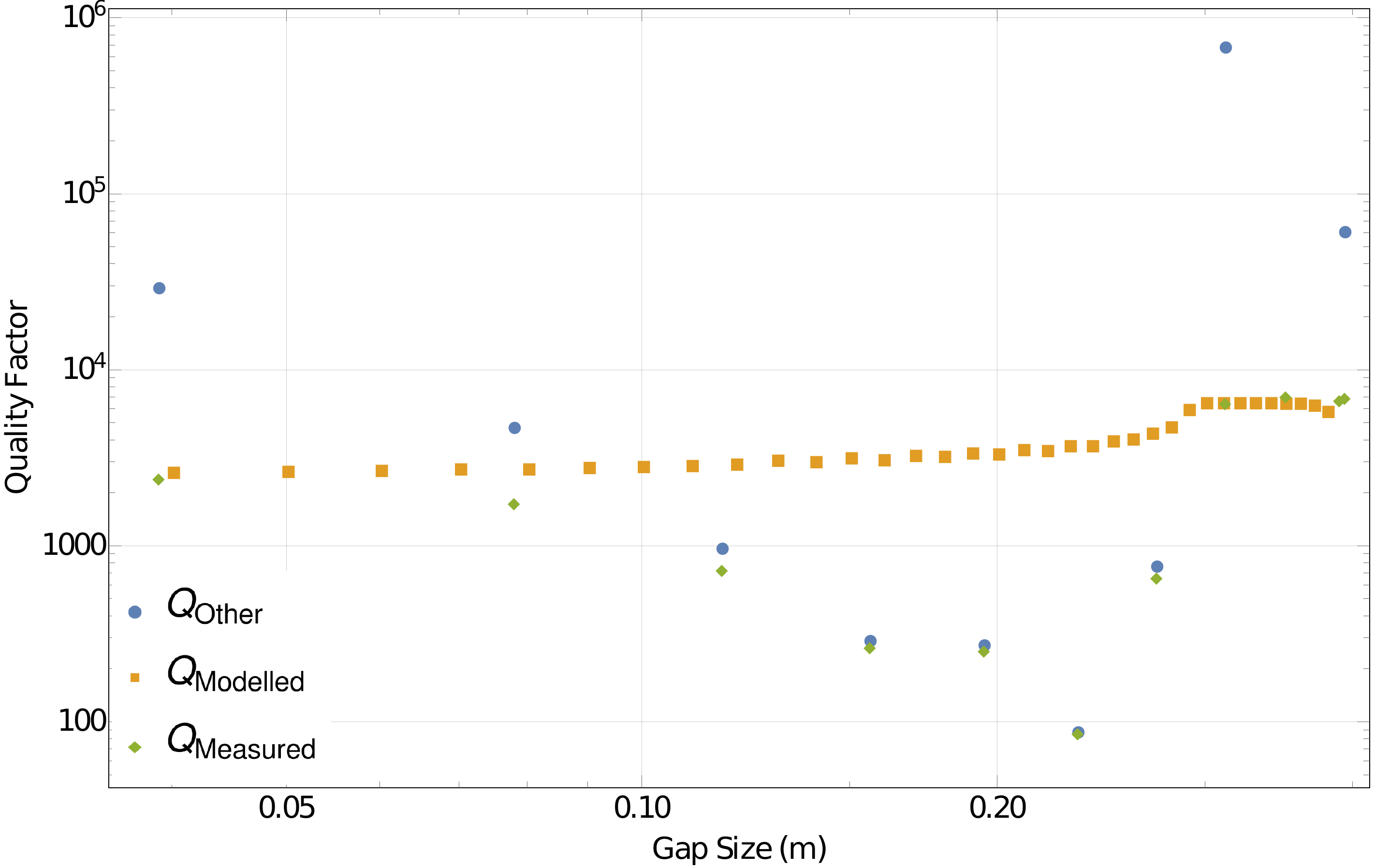}
	\label{fig:Q2}
	\caption{}
\end{subfigure}
	\caption{$Q_{other}$ (blue, circles) computed by comparing measured quality factors (green, diamonds) with quality factors predicted by finite element modelling of geometry factors (yellow, squares) for a) the standard re-entrant mode (left) and b) the first higher order reentrant mode (right). Higher $Q_{Other}$ means lower spurious losses. $Q_{Other}$ values are only computed for points where both measured and simulated data are present, and for which $Q_{Measured}$ is less than $Q_{Modelled}$.}
	\label{fig:Qcombined}
\end{figure*}

The properties of the resonant modes were measured with a vector network analyzer to acquire the complex value of S21 in transmission~\cite{NAQFactor}. Two loop probes were used for the measurement. One probe was located near the base of the central post through a 1-mm-diameter hole in the lid of the cavity, whilst the other was inserted through a hole on the side wall. Both of these probes were used to couple to the $B_\phi$ field component. Fig.~\ref{fig:8} is the device used for the measurement, which consists of an empty cylindrical cavity, a conducting central post, and a micrometer for the movement of the post.
The setup was used to track the frequency of the standard and first higher order mode, and the results are displayed in fig.~\ref{fig:9}. On average, the experimental mode frequencies differed from the theory by less than 3\%.\\
Mode quality factor was also measured via insertion of two weakly coupled magnetic loop probes into the cavity. The measured and predicted unloaded quality factors for each mode are shown in fig.~\ref{fig:Qcombined}. The simulated values assume a conductivity of $2\times10^6$ S/m, which is within the range of values for brass alloys. Brass conductivities are typically on the order of $10^6 - 10^7$ S/m, with variation accounted for by different compositions and manufacturing processes. Initial Q values were significantly lower, this was partially mitigated by polishing the interior surfaces, which resulted in a factor of 2 increase in Q. Further increases in Q were attained by electrically connecting the tuning post to the cavity outer walls. Despite this, there are still losses from a number of factors, including the small gap between the post and the cavity lid, the lack of an RF choke and several other loss mechanisms~\cite{Choke1,Choke2,LowQ1,LowQ2,LowQ3}. We may better model the Q of a real resonator as
\begin{equation}
\frac{1}{Q_0}=\frac{1}{Q_{cav}}+\frac{1}{Q_{\textit{other}}}
\label{eq:Qreal}
\end{equation}
Where $Q_0$ is the measured unloaded quality factor, $Q_{cav}$ is the predicted quality factor from geometry factors as per eq.~\ref{eq:Q}, and $Q_\textit{other}$ is a measure of the other loss mechanisms. We can measure $Q_\textit{other}$ by comparing our computed $Q_{cav}$ with the measured unloaded quality factors. Data for $Q_\textit{other}$ is also presented in fig.~\ref{fig:Qcombined}. It is important to note that there are many possible combinations of $Q_{cav}$ and $Q_\textit{other}$ which could lead to the same value for measured $Q_0$. We present $Q_\textit{other}$ values based on the assumptions outlined above regarding surface resistance of the walls of the resonator. It is indeed possible that the surface resistance is different from that presented and that our findings are in closer agreement, but the pattern  of the measurements indicates some separate loss mechanism that changes as a function of frequency.
We observe good agreement between theory and experiment for small and large gap sizes, however in the case of both modes there is large disagreement for medium gap sizes. We attribute this to some spurious resonant effect, potentially as a result of a highly localized resonance in the small gap between the tuning post and the cavity lid. The high level of agreement between the frequencies found from finite element modelling and the frequencies found experimentally leads us to conclude that we are tracking the correct modes, and that some spurious loss mechanisms, such as gaps around probe holes, and the gap shown in fig.~\ref{fig:8} are responsible.\\ 
\section{Summary and conclusion}
In summary, we have demonstrated the existence of higher order reentrant modes with higher frequencies than the fundamental in the single-post cylrindrical reentrant cavity. We have verified our finite-element modelling by comparison with experimental data, and the agreement is good. We find that the higher order reentrant modes arise from perturbed $TM_{01n}$ modes in empty cylinders, and finally convert to coaxial modes when the gap size becomes zero. The higher order reentrant modes have higher geometry factors when compared to the standard reentrant mode. We have found and characterized the different phases that exist in the transformation process. These modes have possible applications in a number of fundamental physics tests, such as gravity wave and dark matter detection experiments. Application of these modes to an axion haloscope search is discussed, with sensitivity figures of merit presented. Furthermore, it is worth considering that an understanding of these reentrant modes affords us a deeper understanding of the highly localized ``gap modes" present in many experiments. For example, it has been discussed that axion haloscope experiments utilizing tuning rods to adjust mode frequency can have their sensitivity degraded by the presence of highly localized modes which arise due to the small gap between the end of the tuning rod and the walls of the cavity, which must exist for real tuning rods~\cite{AxionGapMode}. 
\begin{acknowledgements}
This work was supported by Australian Research Council grant CE110001013, as well as the UWA-China Research Training Program, the Australian Postgraduate Award and the Bruce and Betty Green Foundation.\\
\end{acknowledgements}
\newpage~\newpage
\begin{appendices}
\section*{Comment on ``Comment on Higher Order Reentrant Post Modes in Cylindrical Cavities [J. Appl. Phys. 122, 144501 (2017)]"}

\section*{abstract}
We recently rigorously described using finite element analysis a cylindrical cavity resonator, with a cylindrical post inserted along the central axis. Such a cavity has a well known reentrant mode where an E$_z$ field exists in the gap between the lid and the post. In J. App. Phys. 122, 144501 (2017)~\cite{Paper} and arXiv:1611.08939 [physics.ins-det] we rigorously analysed higher order modes with similar characteristics to the well known reentrant mode, which we dubbed ``higher order re-entrant post modes". The author claims in arXiv:1801.05418 [physics.app-ph]~\cite{Comment} that these modes have been described before as foreshortened quarter-wave resonators. We discuss the differences between the results of rigorous finite element modelling and the model proposed in the comment, and show that the proposed description is a crude non-Maxwellian model, only approximately valid in a finite region of the cavity tuning range with on average $\sim$9.6\% percent agreement with experimental frequencies for the first higher order mode. The model assumes simple electromagnetic field patterns which do not satisfy Maxwell's equations, and we show that they vary significantly from the rigorous analysis based on Maxwell's equations. The foreshortened quarter-wave resonator model cannot be used to accurately calculate geometry factors, or other factors that require precise knowledge of the fields, such as those computed in the design of cavities for axion experiments. We conclude that the reentrant mode description in the paper is favourable.
\section{Introduction}
The comment arXiv:1801.05418 [physics.app-ph]~\cite{Comment} (``the comment") makes a number of claims regarding the modes discussed in J. App. Phys. 122, 144501 (2017)~\cite{Paper} (``the paper"), as well as a few other claims regarding the discussion and results. We will demonstrate that, whilst the foreshortened quarter-wave resonator model (FQWR model) can generate approximate frequencies across some regions of the tuning range, the predicted field profiles do not reflect the reality of the modes in question, and therefore cannot be used to understand or predict the properties of these modes, which means that the FQWR description is inadequate. We therefore maintain that our rigorous study of these modes using finite element modelling (FEM) is a novel and valuable contribution. It is a crude approximation to refer to these modes as coaxial along the post and capacitive near the gap. Our rigorous FEM study explicitly shows the behaviour of the modes in question, leading us to name them ``higher order reentrant post modes". Whilst similar modes may have been known to the community, we believe they had not been sufficiently, or rigorously studied, and that this new description of the modes as ``higher order reentrant post modes" is more instructive. The fact that the modes are higher order and that there is a sharp edge on the post leads to significant variation from the fields predicted by the FQWR model. Other geometries, such as those described in\cite{Puglisi}, which resemble a rod with a capacitor plate at the end, and without sharp corners may be better suited to modelling with the FQWR model. In particular, the FQWR model is not physically suitable as it does not account for the transition from the small-gap region to the cavity modes, and does not address the reentrant regime, which is the primary interest of the paper.
\section{Frequency Comparison}
The FQWR model in the comment, which comes from~\cite{Puglisi}, is a non-Maxwellian model which supposes that a cavity with a post and a gap between the post and the lid can be adequately represented as a transmission line with standard coaxial modes (the post region), terminated by a capacitor (the gap region). It is claimed that the modes are purely coaxial along the post region, and of ``capacitive character" in the gap. This is only approximately valid in some regions of the tuning range. Particularly the capacitive assumption only holds for very small gaps, and the higher the order the mode is the smaller the gap has to be. The comment presents an equation for finding resonant frequencies of these structures
\begin{equation}
\omega C_0Z_0 - \cot(\frac{2\pi l}{\lambda}) = 0.
\end{equation}
Here $C_0$ is the effective capacitance of the gap region (modelled as a parallel plate capacitor with plate diameter equal to the post diameter), $Z_0$ is the effective impedance of the transmission line region, $l$ is the length of the transmission line region, $\omega$ is the angular frequency of the mode and $\lambda$ the wavelength. Solving this equation yields frequencies that differ from the FEM, and indeed the experiment reported in the paper. Figure 12 shows the frequencies computed with rigorous analysis via FEM, along with the non-Maxwellian FQWR model and the experimental data. The average difference between the FQWR model and the experiment is $\sim$9.6\% for the first higher order mode (if we stop comparing the two at the point where the FQWR model frequency overtakes the next highest frequency mode in the FEM model) and as is apparent the approximation gets worse as the gap size increases. This is unsurprising as the gap region becomes less capacitor-like. We propose that these modes instead be thought of as a reentrant TM modes, as per the paper.
\begin{figure}
\includegraphics[width=\columnwidth]{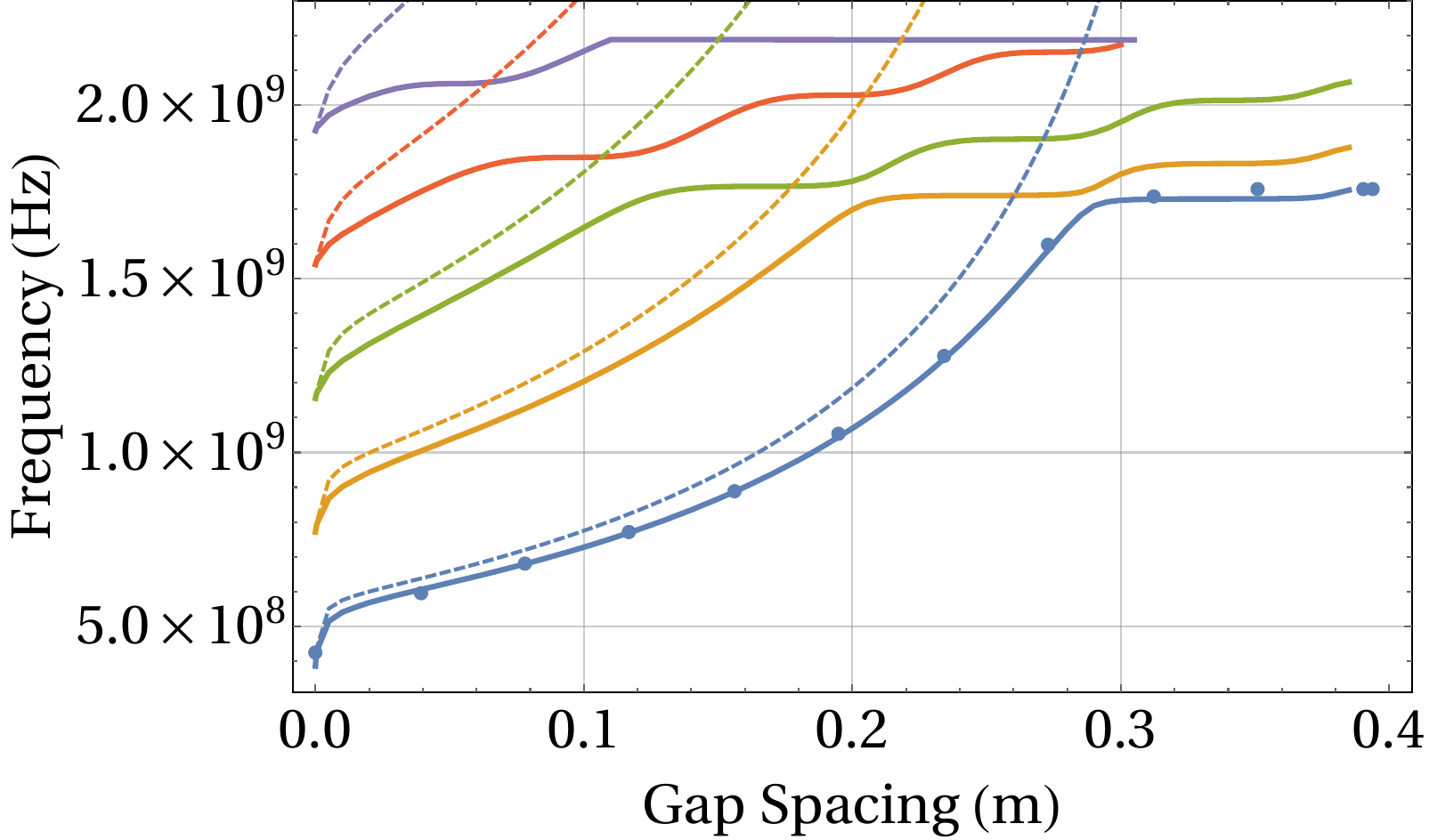}
\caption{Frequencies of the higher order modes in question as a function of gap size computed via FEM (solid lines) and with the FQWR model (dashed lines). The dots represent the data from the experiment in the paper. The fundamental mode is omitted as the present discussion relates to higher order modes.}
\end{figure}
\begin{figure*}
\includegraphics[width=0.92\textwidth]{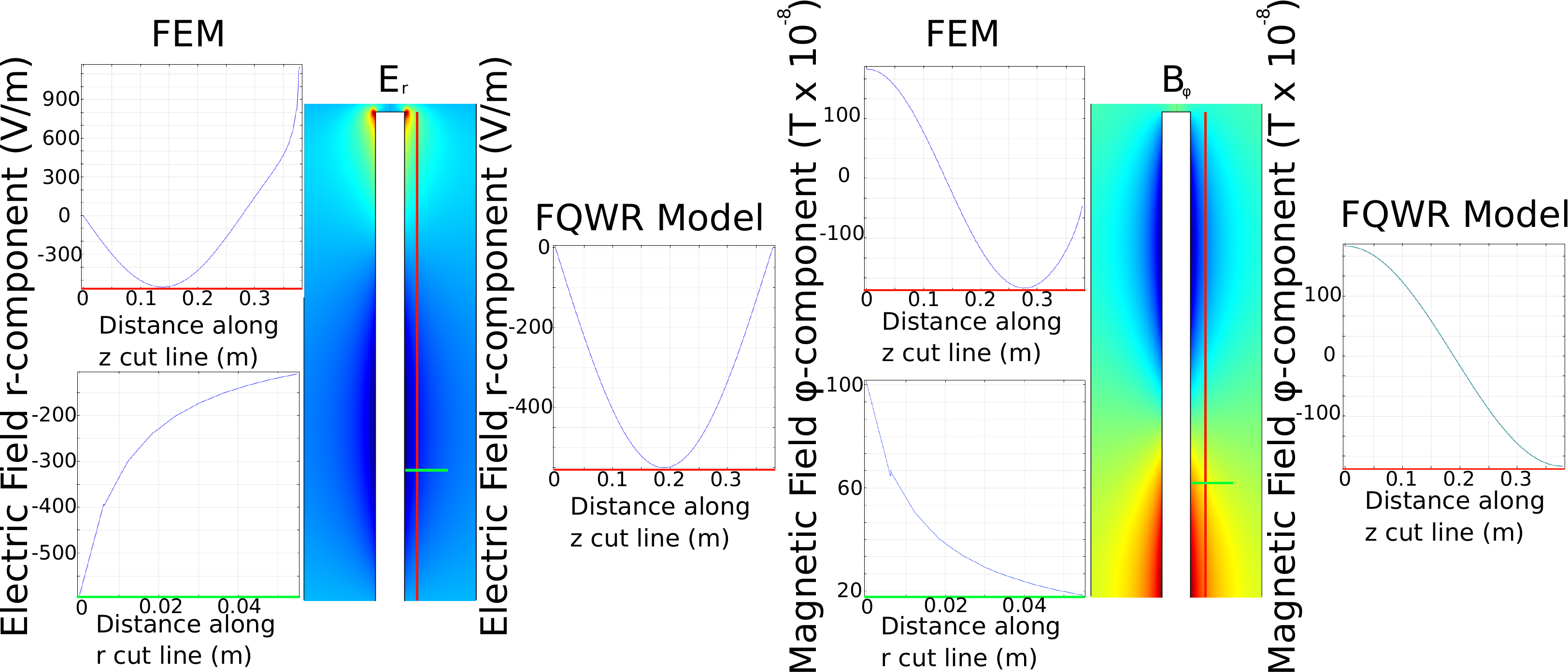}
\caption{Field profiles for the first higher order mode. $E_r$ ($B_\phi$) is shown on the left (right). The graphs on the left show the structure of this field as a function of z and r displacement along the cut lines shown on the figure. The red line represents the z-direction cut, and the field is shown on the top left. The green line represents the r-direction cut, and the field is shown on the bottom left. The field predicted by the FQWR model as a function of z-displacement is shown on the top right of the left (right) hand image.}
\end{figure*} 
%\begin{figure}[t!]
%\includegraphics[width=0.72\columnwidth]{Fig3.pdf}
%\caption{Field profile for the first higher order mode. z-component of electric field is shown. The graphs on the left show the structure of this field as a function of z and r displacement along the cut lines shown on the figure. The blue line represents the z-direction cut, and the field is shown on the top left. The green line represents the r-direction cut, and the field is shown on the bottom right. The FQWR model predicts an $E_z$ field modelled with a simple parallel plate capacitor.}
%\end{figure}
\section{Field Comparison}
In addition to dissimilar frequencies, the FQWR model predicts field profiles which do not reflect properly the reality of the modes. The comment proposes that we can model the fields along the post region as
\begin{align*}
E_r &= E_0\sin(p\pi z/l)e^{-i\omega t}\\
B_\phi &= B_0\cos(p\pi z/l)e^{-i\omega t},
\end{align*}
and states that the electric field in the z-direction should be of ``capacitive character" in the gap region (ie uniform). Rigorous FEM analysis predicts different mode profiles. Fig. 2 shows colour density plots of the first higher order reentrant mode's $B_\phi$ and $E_r$ field components. We show the field profiles as predicted by the FEM, and compare with field profiles from the FQWR model. It was also found that the $E_z$ field component in the gap region is not uniform as a function of z or r, as would be the case for the ``capacitive character" predicted by the FQWR model. Additionally, the model proposed in the comment does not treat the radial profile of the $B_\phi$ or $E_r$ fields, and implicitly assumes that they are uniform as a function of radial position. Figure 13 demonstrates that this is not the case for the FEM model. The FQWR model fails to accurately describe the mode profiles and so is not a useful analytical description, but merely an approximation. We therefore assert that we should not consider these modes simple TEM modes along a post terminated by a capacitor, we should instead view them as higher order reentrant post modes, which arise as perturbed TM modes of empty cavities.\\
If one were to attempt to employ the field profiles from the FQWR model to analytically compute mode dependent quantities such as geometry factors or axion haloscope form factors (as computed in the paper), the results would be incorrect. This further underlines the approximate nature of the FQWR model. The approximate model may have been known to the community for some time, but the reentrant description provides a rigorous and accurate analysis of the mode structures.\\
The FQWR model does not describe the transition of these modes from coaxial to reentrant to TM, as the model becomes increasingly inaccurate at large gaps, and does not account for changing field structures. We can only use the FQWR model to draw approximate conclusions about the behaviour of the fields for very small gaps. For gaps larger than a few hundred microns (with the dimensions modelled in the paper) the field profiles begin to be heavily distorted compared with those derived from the FQWR model. This is critical, as the reentrant regime is the primary region of interest in the paper, and the FQWR model becomes less appropriate further into this regime.\\
It is partially the existence of the sharp edges on the post that create this discrepancy - electric field concentrates at these edges, and distorts the modes from what is described in the FQWR model. Posts with ``softer" corners or plunger-like heads may be better suited to the model, but we maintain that the modes discussed in the paper vary from the simple FQWR model description.
\section{Other Comments}
The comment states: ``While the authors of (the paper in question) recognize the modes as coaxial, they failed to identify them as TEM and make connection to the field in cavities with a non-zero gap". We find this comment unnecessary. We explicitly state that the modes are coaxial at zero gap, and find it unnecessary to state that these coaxial modes are TEM. The modes are no longer TEM for non-zero gaps, and the fact that they are TEM at zero gap has no bearing on this work.\\
Furthermore, it is claimed that we make ``misleading" statements about the relationship between geometry factor and quality factor. In the paper we state that ``the geometry factor ... is directly proportional to the modes quality factor", following promptly with both the full definition of geometry factor, and the explicit relationship between geometry factor and quality factor. We do not agree with the claim that this can be construed as ``misleading", given the immediate presentation of the explicit relationships.
In the comment it is claimed that ``the geometry factor is determined only by the cavity geometry and does not depend on the cavity frequency". This is not quite correct, as for a given cavity geometry different modes will have different geometry factors owing to different mode structures. In a given empty cylindrical cavity the TM$_{020}$ mode will have a different geometry factor than the TM$_{010}$. It is important to examine mode-dependent parameters such as geometry factor, as they allow us to compare different resonant structures before the need to choose a material with a given surface resistance, which does indeed also change with frequency.
\section{Measurement Results}
In criticizing the experimental results presented in the paper, the author of the comment states that ``In addition, the experimental setup described in (the paper in question) lacked proper RF contact (e.g. a sliding spring contact or a choke joint) between the post and the cavity lid. As a result, the measured quality factor values have extremely poor agreement with the calculated values". We detail the experimental setup in the paper, and discuss its limitations due to the lack of a RF choke and the existence of a small gap between the tuning rod and the cavity lid. This comment is unnecessary, given that it restates things stated in the paper. We do not agree that the experimental results are in ``extremely poor" agreement, given that the experimental Qs are within the range of expected values for brass alloys at both very small and large gaps, with disagreement occurring in the middle range of gap spacings, which we attribute to a resonant effect. This, again, is discussed in the paper.
\section{Conclusion}
We discuss a comment on our paper, which claims that the higher order reentrant modes analyzed can be adequately described with a crude non-Maxwellian approximation of the modes as purely coaxial along a post, terminated by a capacitor. We demonstrate that, whilst this model can generate approximate resonant frequencies for some portions of the tuning range, the field structures predicted are inconsistent with those predicted using rigorous finite element modelling. Thus the FQWR model is not a general accurate description, making the reentrant mode description favourable. We discuss the implications of the above, and respond to some other statements made by the author of the comment.
\begin{acknowledgements}
This work was supported by Australian Research Council grant CE170100009, the Australian Postgraduate Award and the Bruce and Betty Green Foundation.\\
\end{acknowledgements}
\end{appendices}

\end{document}